\documentclass{article}

    \PassOptionsToPackage{numbers, compress}{natbib}

\usepackage[preprint]{neurips_2022}

\usepackage[utf8]{inputenc} 
\usepackage[T1]{fontenc}    
\usepackage{hyperref}       
\usepackage{url}            
\usepackage{booktabs}       
\usepackage{amsfonts}       
\usepackage{nicefrac}       
\usepackage{microtype}      
\usepackage{xcolor}         
\usepackage{subcaption}
\usepackage{natbib}
\usepackage{hyperref}
\hypersetup{
    colorlinks=true,
    citecolor=black,
    linkcolor=red,
    urlcolor=cyan
}

\title{Growing Brains: Co-emergence of Anatomical and Functional Modularity in Recurrent Neural Networks}

\author{%
  Ziming Liu\thanks{Equal contribution} \\
  MIT \& IAIFI \\
  \texttt{zmliu@mit.edu} \\
  \And
  Mikail Khona$^*$ \\
  MIT \\
  \texttt{mikail@mit.edu} \\
  \AND
  Ila R. Fiete \\
  MIT \\
  \texttt{fiete@mit.edu} \\
  \And
  Max Tegmark \\
  MIT \& IAIFI \\
  \texttt{tegmark@mit.edu} \\
}

\usepackage{graphicx}
\usepackage{dcolumn}
\usepackage{bm}
\usepackage{float}
\usepackage{subcaption}
\usepackage{multirow}
\usepackage{comment}
\usepackage{dsfont}
\usepackage{amsthm}
\usepackage{xcolor}
\usepackage{array}
\usepackage{eucal}
\usepackage{makecell}
\usepackage{mathtools}
\usepackage{makecell}
\newcommand{\yangtasks}{\texttt{20-Cog-tasks} }
\newcommand{\modcogtasks}{\texttt{Mod-Cog-tasks} }
\usepackage{enumitem}
\setlist{leftmargin=10mm}

\newcommand{\mat}[1]{\mathbf{#1}}

\def\spose#1{\hbox to 0pt{#1\hss}}
\def\simlt{\mathrel{\spose{\lower 3pt\hbox{$\mathchar"218$}}
     \raise 2.0pt\hbox{$\mathchar"13C$}}}
\def\simgt{\mathrel{\spose{\lower 3pt\hbox{$\mathchar"218$}}
     \raise 2.0pt\hbox{$\mathchar"13E$}}}
\def\simpropto{\mathrel{\spose{\lower 3pt\hbox{$\mathchar"218$}}
     \raise 2.0pt\hbox{$\propto$}}}

\def\beq#1{\begin{equation}\label{#1}}
\def\eeq{\end{equation}}
\def\beqa#1{\begin{eqnarray}\label{#1}}
\def\eeqa{\end{eqnarray}}

\begin{document}

\maketitle

\begin{abstract}
Recurrent neural networks (RNNs) trained on compositional tasks can exhibit functional modularity \cite{yang2019task, khona2023winning}, in which neurons can be clustered by activity similarity and participation in shared computational subtasks. Unlike brains, these RNNs do not exhibit \emph{anatomical modularity}, in which functional clustering is correlated with strong recurrent coupling and spatial localization of functional clusters. Contrasting with functional modularity, which can be ephemerally dependent on the input \cite{khona2023winning}, anatomically modular networks form a robust substrate for solving the same subtasks in the future. To examine whether it is possible to grow brain-like anatomical modularity, we apply a recent machine learning method, brain-inspired modular training (BIMT), to a network being trained to solve a set of compositional cognitive tasks. We find that functional and anatomical clustering emerge together, such that functionally similar neurons also become spatially localized and interconnected. Moreover, compared to standard $L_1$ or no regularization settings, the model exhibits superior performance by optimally balancing task performance and network sparsity. In addition to achieving brain-like organization in RNNs, our findings also suggest that BIMT holds promise for applications in neuromorphic computing and enhancing the interpretability of neural network architectures.
\end{abstract}




\section{Introduction}

A powerful way for networks to generalize is through modularity: If seen and unseen tasks in the world consist of combinations of  subtasks, then a new task can be quickly solved by decomposing it into the set of previously seen subtasks, and tackling those based on prior learning. Recent work shows that RNNs trained on a set of tasks drawn by combining subtasks from a common dictionary begin to exhibit functional modularity, with similar activity profiles across neurons responding to the same subtask. However, the formed clusters were not anatomical. Anatomical clustering with localization of function is a central feature of brains \cite{ferrier1874localization}: for example, visual processing for object recognition is localized to the ventral visual pathway while the initiation of voluntary movements is confined to a few motor and premotor cortical regions. Anatomical modularization can facilitate continual learning: if it takes the form of spatial localization, then new inputs can be easily routed to the module; if it takes the form of recurrent connectivity, it provides lasting substructures for solving specific computations on future tasks. By contrast, functional clustering alone can be ephemeral, with groupings that might be defined primarily by correlations in the inputs \cite{khona2023winning}. When the input correlations change, functional modules could disappear. 

Recently, the method of brain-inspired modular training (BIMT) was proposed as a way to make artificial neural networks modular and more interpretable~\cite{liu2023seeing}. The key idea of BIMT is to encourage local neural connections via two optimization terms: distance-dependent weight regularization and discrete neuron swapping. Here we ask whether BIMT can answer a fundamental question about neuroscience: Can spatial constraints and wiring costs ~\cite{chen2006wiring,chklovskii2002wiring,chklovskii2004maps} together lead to the emergence of anatomical modules that are also functionally distinct? 

We study how BIMT can lead to the emergence of spatial modules in a multitask learning setting relevant to cognitive systems neuroscience with two sets of combinatorially constructed tasks, \yangtasks \cite{yang2019task} and \modcogtasks \cite{khona2023winning}. We train recurrent neural networks (RNNs) on these tasks with BIMT in the supervised setup. We observe brain-like spatial organization emerging in the hidden layer of the RNN: neurons that are functionally similar are also localized in space (Figure~\ref{fig:basic-results-2D}c). Such locality and sparsity are gained with no sacrifice in performance, or even accompanied by an improvement in performance. We introduce our methods in Section~\ref{sec:method}, and present results in Section~\ref{sec:results}. Due to limited space, the main paper focus on \yangtasks \cite{yang2019task} and leave the results for \modcogtasks \cite{khona2023winning} in Appendix~\ref{app:1D-84}.

\section{Method}\label{sec:method}

\begin{figure}[tbp]
    \centering
    \includegraphics[width=1.0\linewidth]{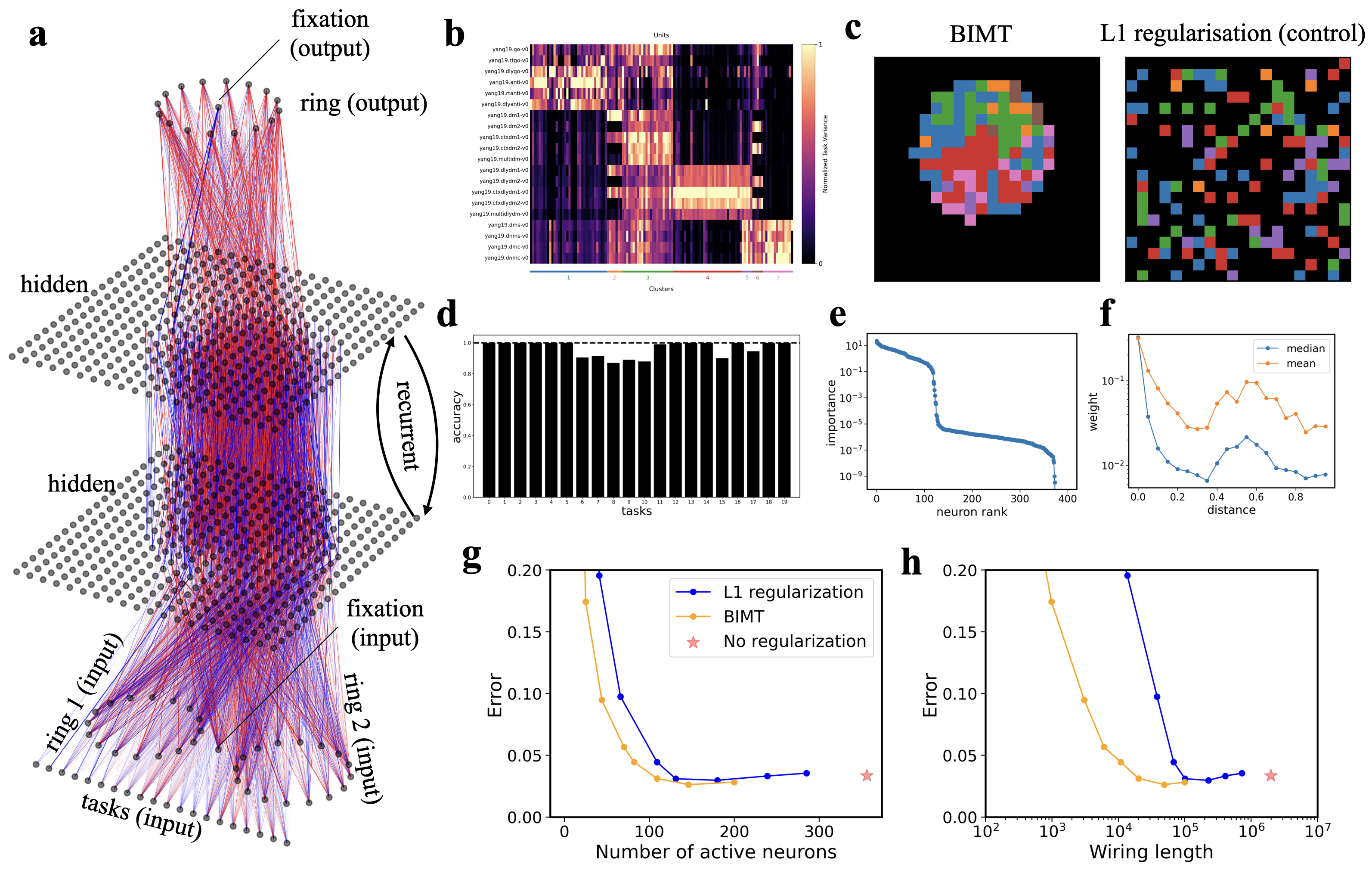}
    \caption{Training an RNN with BIMT on cognitive tasks. {\bf a}: Visualization of the network. Each line represents a weight; blue/red means positive/negative weights; thickness corresponds to magnitudes. {\bf b}: The hidden neurons are clustered into functional modules. {\bf c}: These functional modules (distinguished by colors) are also clustered in space, visually resembling a brain. By contrast, $L_1$ regularization leads to no anatomical modules. The network has ({\bf d}) good performance, ({\bf e}) high sparsity and good locality ({\bf f}). {\bf g} and {\bf h}: Trade-off between performance (error) and sparsity. {\bf g} uses the number of active neurons as the sparsity measure, while {\bf h} uses the wiring length as the sparsity measure. In both cases, the Pareto frontier of BIMT is better than that of $L_1$ regularization and no regularization.}
    \label{fig:basic-results-2D}
\end{figure}

{\bf Brain-inspired modular training (BIMT)} Biological neural networks (e.g., brains) differ from artificial ones in that biological ones restrain neuronal connections to be local in space, leading to anatomical modularity. Motivated by this observation,~\cite{liu2023seeing} proposed brain-inspired modular training (BIMT) to facilitate modularity and interpretability of artificial neural networks. The idea is to embed neurons into a geometric space and minimize the total connection cost by adding the connection cost as a   penalty to the loss function and swapping neurons if necessary. On a number of math and machine learning datasets, they show that BIMT is able to put functionally relevant neurons close to each other in space, just like brains. It is thus natural to ask: Can BIMT give something back to neuroscience? In this work, we apply BIMT to recurrent neural networks (RNN) for cognitive tasks, and show that the neurons in the hidden layer are organized into modules which are both anatomically and functionally distinct, just like brains.

{\bf RNN} We take a simple recurrent neural network (RNN) in the context of systems neuroscience, which is defined by 
\begin{equation}
    \begin{aligned}
        & \mat{h}_{t+1} = \phi(\mat{W}\mat{h}_t + \mat{W}_{\rm in}\mat{u}_t+\mat{b}^h), \\
        & \mat{o}_{t+1} = \mat{W}_{\rm out}\mat{h}_{t+1} + \mat{b}^o,
    \end{aligned}
\end{equation}
where $\mat{u}_t\in\mathbb{R}^{n_u}$, $\mat{h}_t\in\mathbb{R}^{n_h\times n_h}$, $\mat{o}_t\in\mathbb{R}^{n_o}$, $\mat{W}_{\rm}\in\mathbb{R}^{n_h\times n_h}\times \mathbb{R}^{n_h\times n_h}$. We place $n_h\times n_h$ hidden neurons uniformly on a 2D grid $[0,1]^2$ (see Figure~\ref{fig:basic-results-2D}a), so the $ij$ neuron (the neuron in the $i^{\rm th}$ row and the $j^{\rm th}$ column) is located at $(i/n_h,j/n_h)$. The $L_1$ distance between the $ij$ neuron and the $mn$ neuron is thus $\mat{D}_{ij,mn}\equiv (|i-m|+|j-n|)/n_h$~\footnote{One can choose other distances, e.g., $L_2$ distance. We choose $L_1$ distance because it is more consistent with $L_1$ regularization.}. We define the RNN's connection cost as 
\begin{equation}
    \ell_{cc} = \underbrace{\|\mat{W}\|_1+\|\mat{W}_{\rm in}\|_1+\|\mat{W}_{\rm out}\|_1+|\mat{b}^h|_1+|\mat{b}^o|_1}_\text{vanilla $L_1$ regularization} + \underbrace{A\|\mat{D}\odot\mat{W}\|_1}_\text{distance-aware regularization},
\end{equation}
where $\|\mat{M}\|_1\equiv \sum_{ij} |M_{ij}|$ and $|\mat{v}|_1\equiv \sum_{i} |v_{i}|$ are matrix and vector $L_1$-norm, respectively. $A$ is a hyper-parameter controling the strength of locality constraint.

{\bf Cognitive tasks} The \yangtasks are a set of simple cognitive tasks inspired by experiments with rodents and non-human primates performed by systems neuroscientists \cite{yang2019task}. These tasks are designed to fall into families where each family is defined by a set of computations drawn from a common pool of computational primitives. Thus, the tasks have shared subtasks and an optimal solution is to form clusters of neurons specialized to these subtasks and share them across tasks, illustrated in Figure~\ref{fig:hist_ood_2D}a. In our experiments, we set both input rings to have 16 dimensions, so input dimension $n_u = 16\times 2 + 20 + 1 = 53$ (where 20 for the one-hot length-20 task vector, and 1 for fixation). Hidden neurons are aranged as $20\times 20$ grid ($n_h=20$), and output dimension $n_o=16+1=17$. The prediction loss $\ell_{\rm pred}$ is the cross-entropy between the ground truth and the predicted reaction.

{\bf BIMT} loss simply combines the prediction loss and the connection cost, i.e., the total loss function is
\begin{equation}
    \ell = \ell_{\rm pred} + \lambda\ell_{cc},
\end{equation}
where $\lambda\geq 0$ is the strength of penalizing connection costs. When $\lambda=0$, it boils down to train a fully-connected RNN without sparsity constraint; when $\lambda>0$ but $A=0$, it boils down to train with vanilla $L_1$ regularization. Besides adding connection costs as regularization, BIMT allows swapping neurons to further reduce $\ell_{cc}$ by avoiding local minima, e.g., if a neural network is initialized to be performing well but has non-local connections, without swapping, the network would not change much and maintain those non-local connections. 

\section{Results}\label{sec:results}

We focus first on results in the \yangtasks; qualitatively similar results on the \modcogtasks are included in Appendix~\ref{app:1D-84}. 

\subsection{BIMT learns a 2D brain that solves all \yangtasks}

We show results for $A=1.0$ and $\lambda=10^{-5}$ in Figure~\ref{fig:basic-results-2D}. {\bf a} shows the connectivity graph of the RNN. All the weights are plotted as lines whose thicknesses are proportional to their magnitudes~\footnote{Some weights are visually vanishing, because their magnitudes are too small, although we indeed plot them.}, and blue/red means positive/negative weights. BIMT learns to prune away peripheral neurons and concentrate important neurons only in the middle. Following~\cite{yang2019task}, we cluster neurons into functional modules based on their normalized task variance (shown in {\bf b}). These functional modules, colored by different colors, are shown to be also anatomically modular, i.e., spatially local (shown in {\bf c}), visually resembling a brain. By contrast, $L_1$ regularization does not induce any anatomical module. {\bf d} shows that the network performs reasonably well. The network is sparse; {\bf e} shows that it only contains around 100 important neurons (measured by sum of task variances). The connections in the hidden layer are mostly local, as we hoped; {\bf f} shows that weights decay fast as distance increases, which is similar to fixed local-masked RNN~\cite{khona2023winning}. However intriguingly, there is a second peak of (relatively) strong connections around distances being 0.6, which is probably attributed to inter-module connections.

\subsection{Sparsity vs Accuracy Tradeoff}

There is a Pareto frontier showing the trade-off between sparsity and accuracy, shown in Figure~\ref{fig:basic-results-2D}{\bf g} and {\bf h}. We also compute the trade-off for networks with vanilla $L_1$ regularization. We use two sparsity measures: the number of active neurons (a neuron is active if its sum of tasks variances is larger than $10^{-3}$), and the wiring length (sum of lengths of all active connections; a connection is active if its weight magnitude is larger than $10^{-2}$). Under both measures, BIMT is superior than $L_1$ regularization in terms of having better Pareto frontier.

\subsection{Anatomical modularity}

\begin{figure}[tbp]
    \centering
    \includegraphics[width=1.0\linewidth]{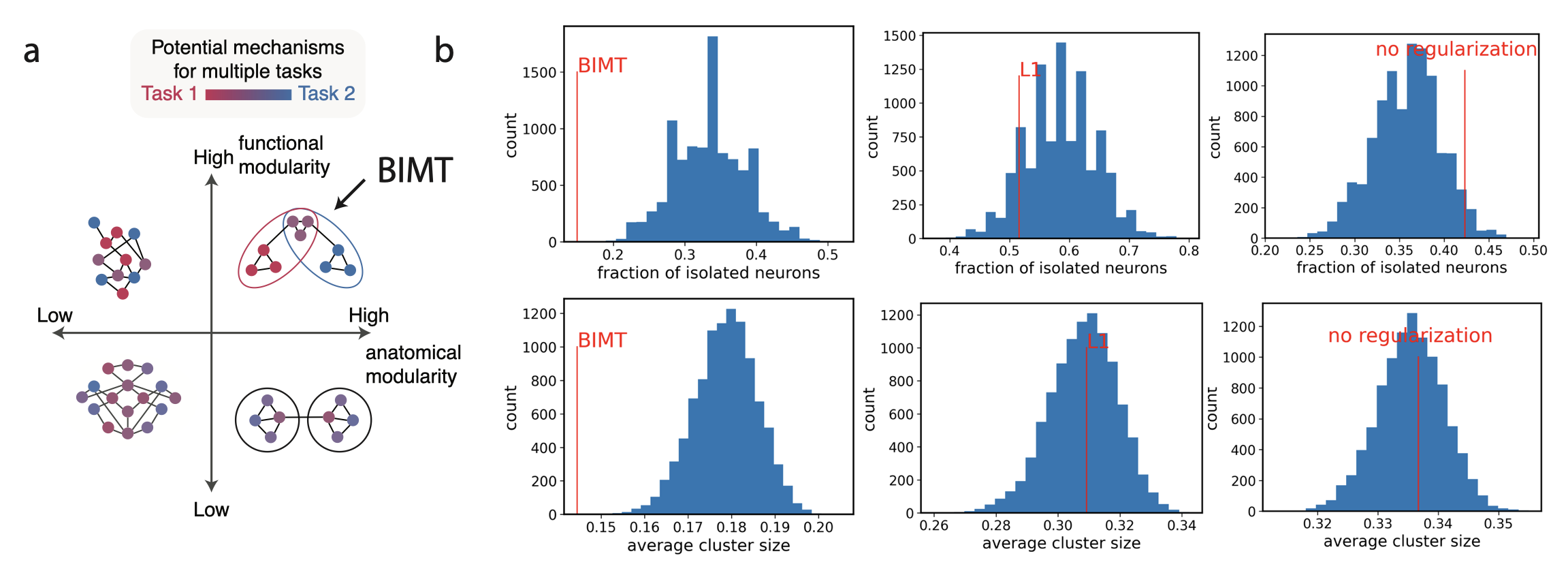}
    \caption{(a) The myriad ways anatomical and functional modularity can present itself in trained RNNs. (b) We test anatomical modularity for neural networks with BIMT (left), L1 regularization (middle) or no regularization (right). We propose two metrics, fraction of isolated neurons (top) and average (functional) cluster size (bottom) to measure anatomical modularity. For both metrics, smaller is better. We compare the trained network with networks whose useful hidden neurons are randomly shuffled. No regularization and L1 regularization are in the distributions of randomly shuffled networks, while BIMT is significantly out of distribution (smaller) than random ones, indicating anatomical modularity.}
\label{fig:hist_ood_2D}
\end{figure}

Anatomical modularity means that neurons with similar functions are placed close to each other in space. Because neurons of fully-connected layers have permutation symmetries, there is no incentive for them to develop anatomical modularity. By contrast, since BIMT penalizes connection costs, BIMT networks potentially have anatomical modularity. In Figure~\ref{fig:basic-results-2D}{\bf c}, each neuron's functional cluster is marked and there are clear spatial clusters in which all neurons belong to the same functional cluster. Quantitatively, we propose two metrics: (1) the fraction of isolated neurons. A neuron is isolated if none of its (eight) neighbors belongs to the same functional cluster. (2) the average size of functional clusters. For both metrics, the smaller the better. For baselines, we randomly shuffle important hidden neurons~\footnote{A neuron is important if the sum of its task variances is above $10^{-3}$.}. Since different random shuffling may yield different results, we try 10000 different random seeds and plot histograms for the metrics. We compute the two metrics for networks trained with BIMT, $L_1$ regularization, or no regularization in Figure~\ref{fig:hist_ood_2D}. Only BIMT networks are seen to be significantly out-of-distribution from baselines, implying anatomical modularity. In future work, we hope to explore improvements of BIMT to further increase the functional modularity of the ``brain" seen in Figure~\ref{fig:basic-results-2D}{\bf c}.

Our main findings remain qualitatively similar when (i) the topology of hidden layer is changed or (ii) the tasks are significantly harder, so that they require more-involved recurrent connectivity within the network. 
Specifically, in Appendix~\ref{app:1D-20}, we include the results for 1D hidden layer. In Appendix~\ref{app:1D-84}, we find that results on a more-complex set of 84 cognitive tasks, the \modcogtasks~\cite{khona2023winning}, show qualitatively similar anatomical clustering results, demonstrating the robustness and generality of our core findings.

\section*{Acknowledgement}
ZL and MT are supported by IAIFI through NSF grant PHY-2019786, the Foundational Questions Institute and the Rothberg Family Fund for Cognitive Science. IRF is supported by the Simons Foundation through the Simons Collaboration on the Global Brain, the ONR, the Howard Hughes Medical Institute through the Faculty Scholars Program and the K. Lisa Yang ICoN Center. MK acknowledges funding from the Department of Physics, MIT.

\bibliography{BIMT_cog.bib}

\newpage
\appendix
{\huge Appendix}

\section{1D results for \yangtasks}\label{app:1D-20}

\begin{figure}[htbp]
    \centering
    \includegraphics[width=1\linewidth]{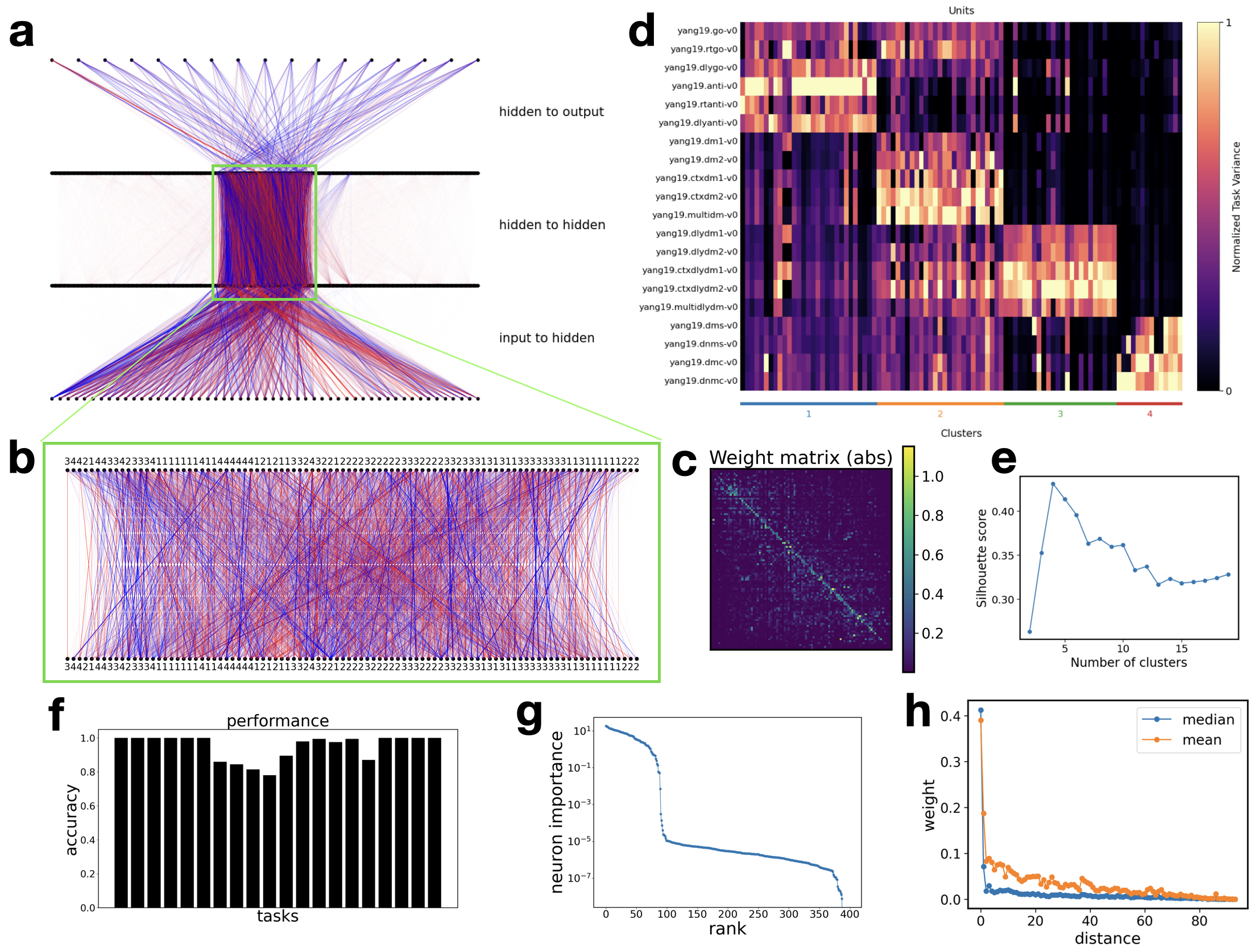}
    \caption{Training an RNN with BIMT on cognitive tasks. {\bf a}: Visualization of the network. Each line represents a weight; blue/red means positive/negative weights; thickness corresponds to magnitudes. The hidden-to-hidden connections are zoomed-in in {\bf b}, and displayed as a weight matrix in {\bf c}, which is strongly diagonal. {\bf d} and {\bf e}: We decompose important neurons into  clusters based on task variance~\cite{yang2019task}. The BIMT  network has reasonable performance ({\bf f}), high sparsity ({\bf g}) and high locality ({\bf h}).}
    \label{fig:basic-results}
\end{figure}

\begin{figure}[htbp]
    \centering
    \begin{subfigure}[]{0.32\textwidth}
    \includegraphics[width=\linewidth, trim=0cm 0cm 0cm 0cm]{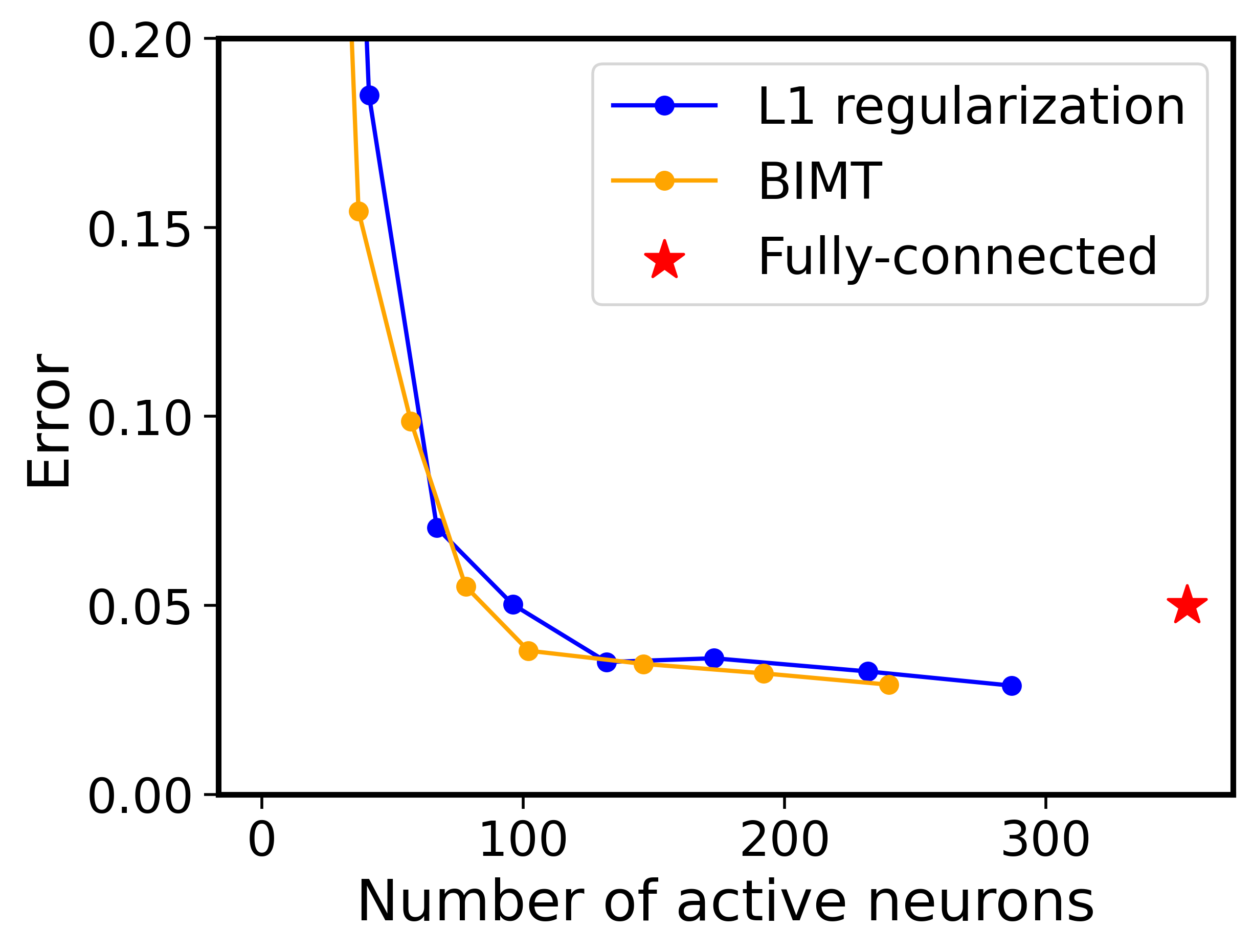}
    \caption{}
    \label{fig:compare}
    \end{subfigure}
    \begin{subfigure}[]{0.32\textwidth}
    \includegraphics[width=\linewidth, trim=0cm 0cm 0cm 0cm]{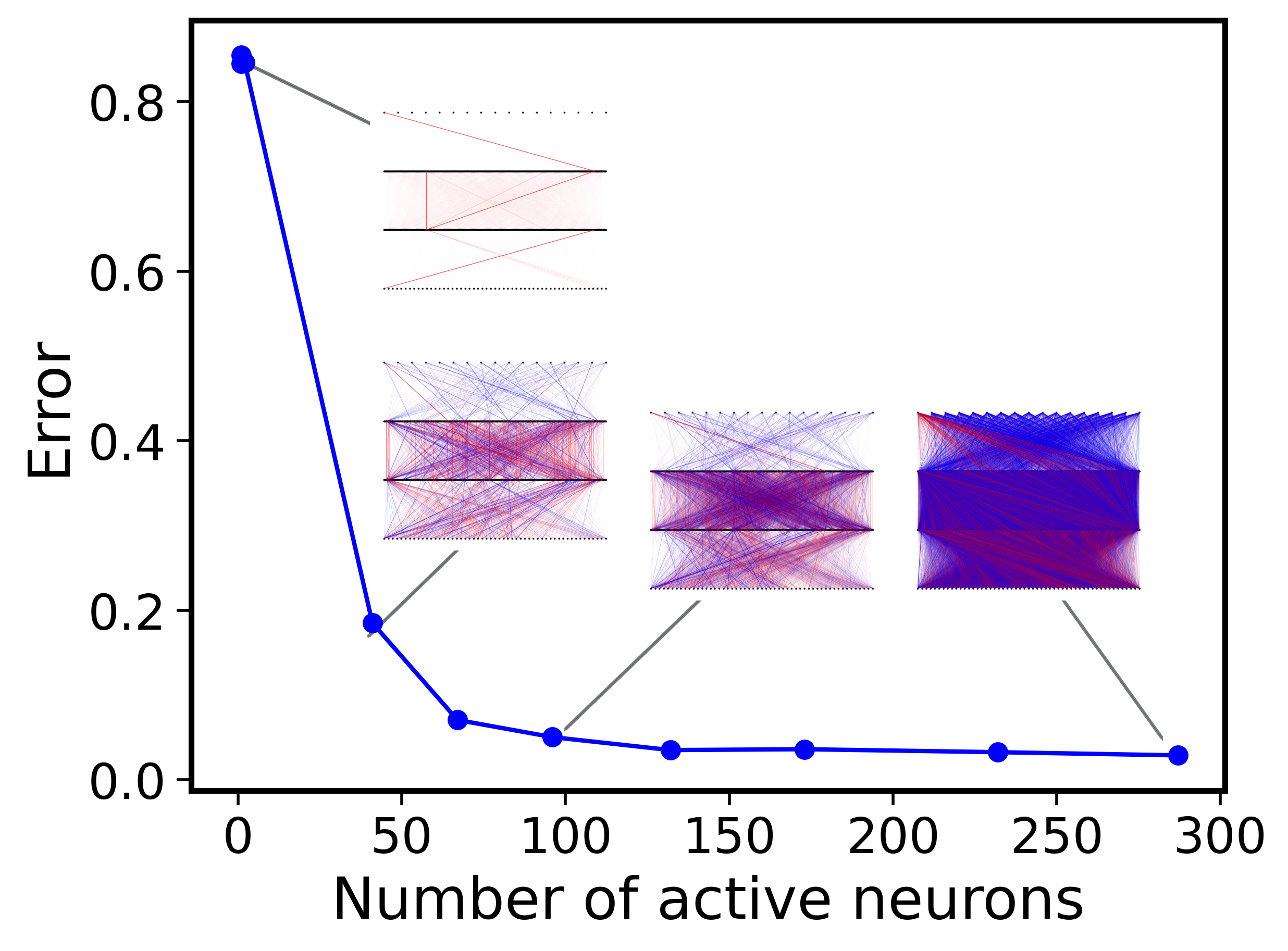}
    \caption{}
    \label{fig:L1}
    \end{subfigure}
    \begin{subfigure}[]{0.32\textwidth}
    \includegraphics[width=\linewidth, trim=0cm 0cm 0cm 0cm]{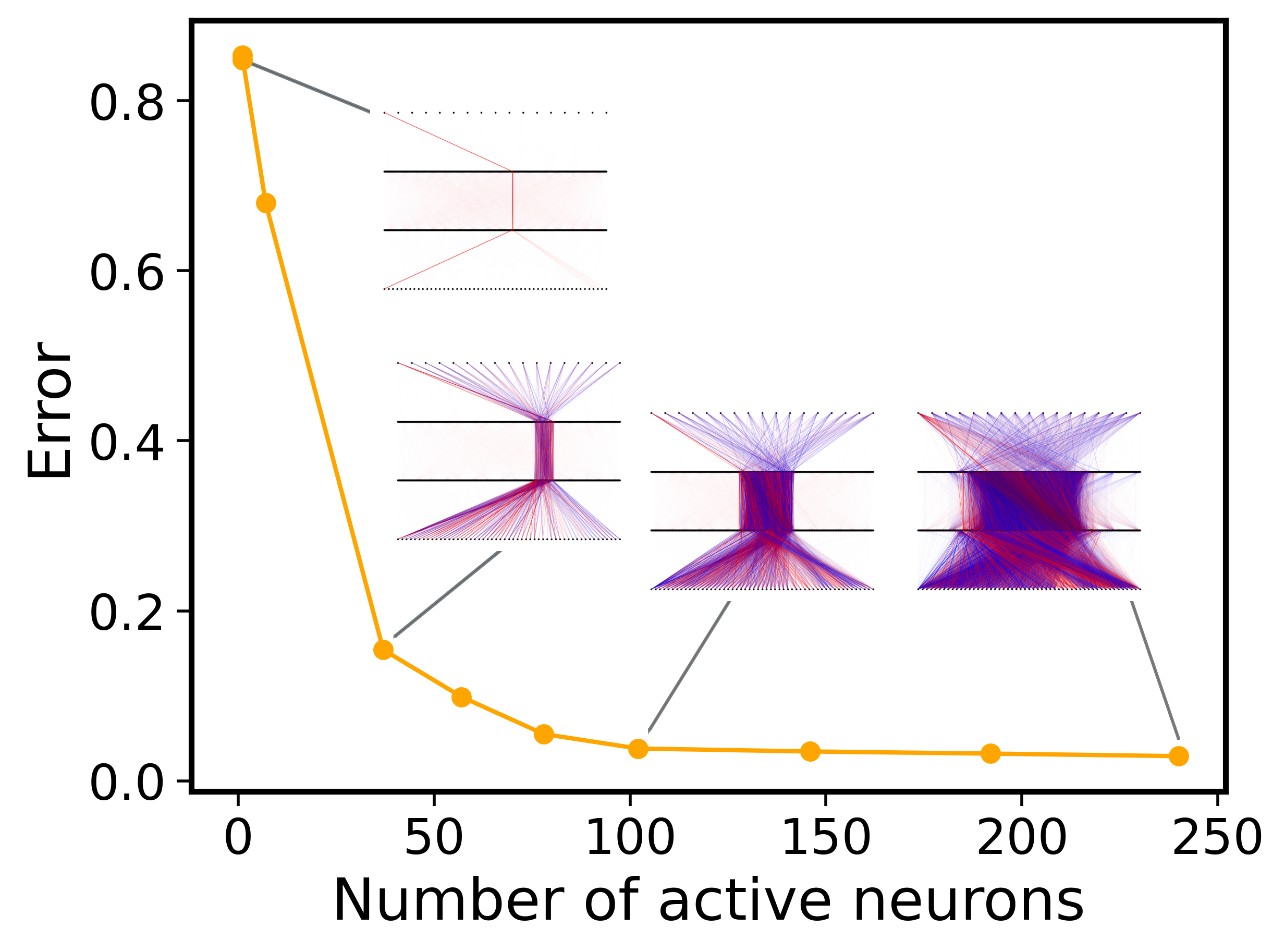}
    \caption{}
\label{fig:BIMT}
    \end{subfigure}
    \caption{Sparsity-accuracy tradeoff and spatial structures for neural networks. (a) BIMT has a similar Pareto frontier to the baseline L1 regularization, while fully-connected network is neither sparse nor optimally performant. (b) Networks trained with $L_1$ penalty do not have spatial structures, while (c) BIMT networks learn to prune away peripheral hidden  neurons, and have anatomical modularity (see Figure~\ref{fig:hist_ood}).}
    \label{fig:sparsity-acc-tradeoff}
\end{figure}

\begin{figure}[htbp]
    \centering
    \includegraphics[width=1\linewidth]{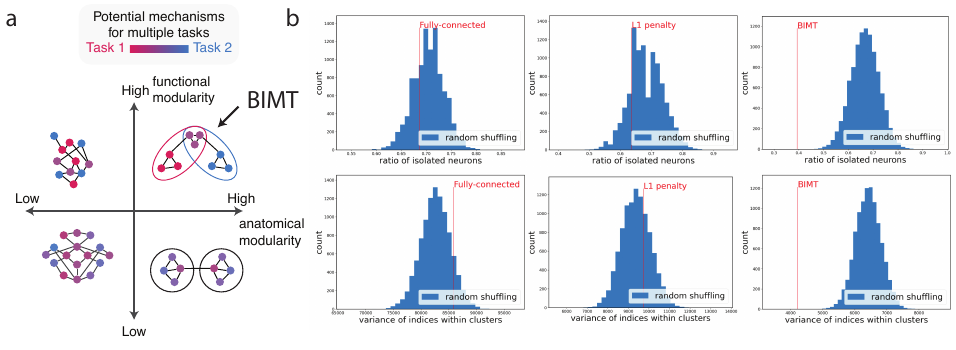}
    \caption{(a) The myriad ways anatomical and functional modularity can present itself in trained RNNs. (b) We test anatomical modularity for neural networks with vanilla training (left), L1 penalty (middle) or BIMT (right). We propose two metrics, ratio of isolated neurons (top) and variance of indices within clusters (bottom) to measure anatomical modularity. For both metrics, smaller is better. We compare the trained network with networks whose hidden neurons are randomly shuffled. Fully-connected networks and L1 networks are in the distributions of randomly shuffled networks, while BIMT is significantly out of distribution (smaller) than random ones, indicating anatomical modularity.}
    \label{fig:hist_ood}
\end{figure}

In the main paper, we lay hidden neurons on a 2D grid, which is mainly because the neocortex is effectively a two-dimensional sheet. Algorithm-wise, BIMT allows any geometric space. For example, we can arrange hidden neurons into a one-dimensional grid. Results are included in Figure~\ref{fig:basic-results},\ref{fig:sparsity-acc-tradeoff} and \ref{fig:hist_ood}. 

\section{Results for 84 \modcogtasks}\label{app:1D-84}
\begin{figure}
    \centering
    \includegraphics[width=1\linewidth]{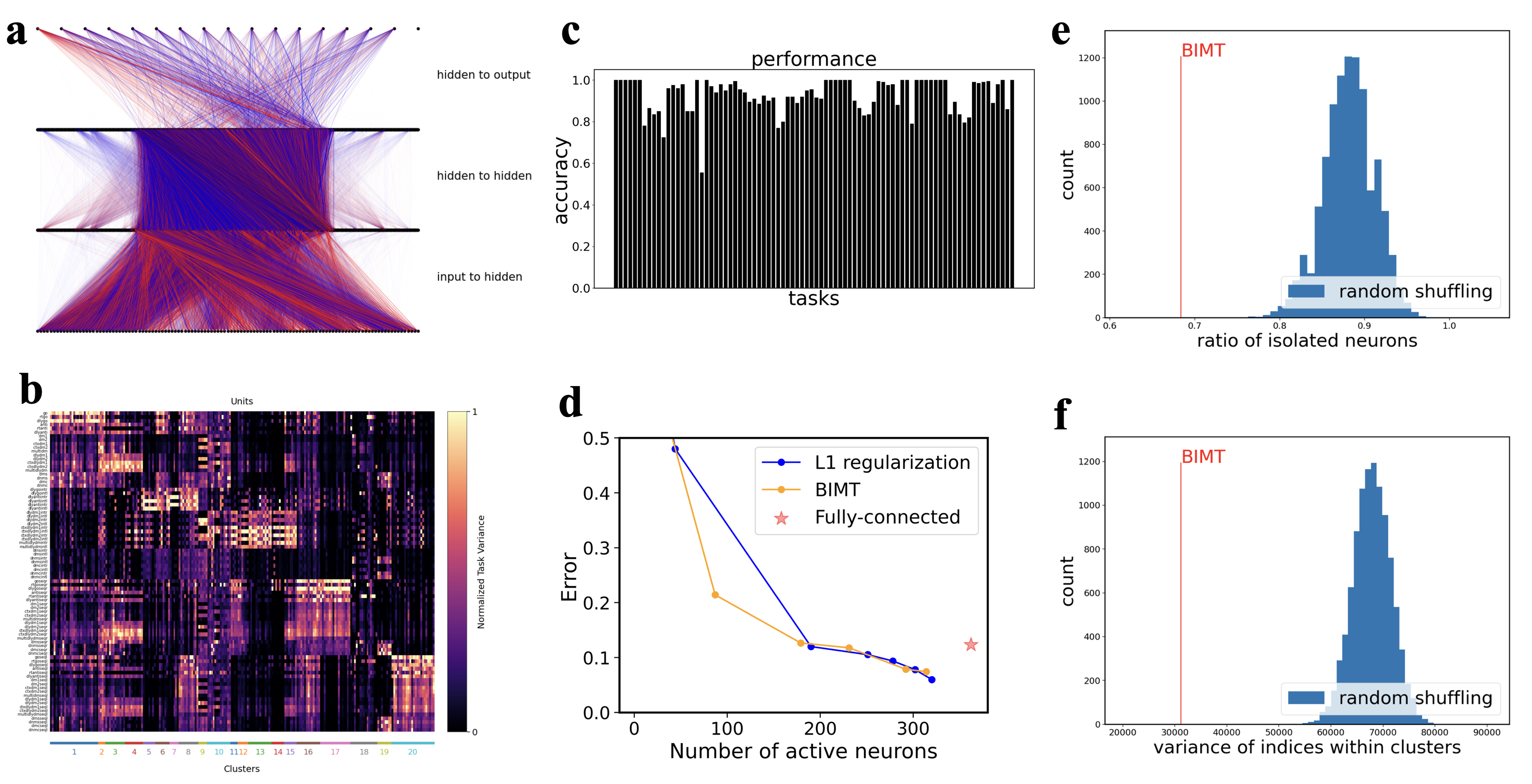}
    \caption{Results for 84 Mod-Cog tasks. {\bf a}: the visualization of the connectivity graph of the 1D RNN; {\bf b}: neurons are grouped into functional modules based on task variances; {\bf c}: the network has high performance for most tasks; {\bf d}: the trade-off between performance and sparsity. The pareto frontier of BIMT is similar to that of $L_1$, which means that anatomical modularity can be achieved without sacrifice in performance or sparsity. {\bf e} and {\bf f}: RNNs trained with BIMT clearly demonstrate anatomical modularity since the two metrics (fraction of isolated neurons and variance of indices within clusters) are significantly out-of-distribution of random baselines. }
    \label{fig:mod-cog}
\end{figure}

The \yangtasks are a relatively simple set that can be solved with only autapses in the hidden layers \cite{khona2023winning}, raising the question of whether, when subtasks require extensive recurrent connectivity in the hidden networks, BIMT will lead to emergence of anatomical modularity. Khona et al. \cite{khona2023winning} extended Yang 20 tasks~\cite{yang2019task} to a more diverse set of tasks (84 in total) that require recurrent connectivity among multiple neurons within each module. We find that the conclusions in the main paper also hold for this augmented suite of cognitive tasks, demonstrating the robustness of our claims. Results are included in Figure~\ref{fig:mod-cog}.

\end{document}